# Enhancement of the Rate of Surface Reactions by Elasto-capillary Effect

Nitish Singh[1] and Animangsu Ghatak[1,2,*]

[1] Department of Chemical Engineering, Indian Institute of Technology Kanpur, 208016
[2] Center for Environmental Science and Engineering, Indian Institute of Technology Kanpur, 208016 (India)
[*] Prof. A. Ghatak, Corresponding-Author, Author-Two, E-mail: aghatak@iitk.ac.in

**Abstract:**

Rate of a reaction is enhanced by increase in temperature, partial pressure or the concentration of reactants, and/or via use of a catalyst or an enzyme. Whereas the former increases the probability of collision rate between different molecules, the latter provides an alternate reaction path that diminishes the activation energy barrier for reaction, both effects increase the reaction rate. The deformability of the substrate on which a reaction occurs is not known to affect this rate. In contrast, by carrying out reduction of Gold and Silver salt on soft, crosslinked layers of poly(dimethylsiloxane) (PDMS), we show here that surface tension driven deformation of the solid too can increase the reaction rate. The PDMS contains SiH groups as organosilanes which reduces the salt thereby producing the corresponding metallic nano-particles. When an aqueous solution of the salt is dispensed as a sessile drop on a sufficiently soft PDMS layer, nano-particles get generated at ~3 times the rate over which when the salt solution is dispensed as a pool on the same surface. Unlike the pool, the drop has a three-phase contact line, where it pulls up the solid forming a ridge. By balancing the resultant surface stresses at the vicinity of this ridge, we have estimated the excess surface free energy associated with it. We show that this excess energy can diminish the activation energy barrier and increase the reaction rate to the extent that would have been achieved by increasing the reaction temperature by over 30º C.

**Significance Statement:**

While many applications demand the rate of a reaction to be enhanced, parameters which are varied to accomplish it, are temperature, partial pressure or concentration of reactants and a catalyst. We show here show that deformability of a soft solid, that hosts such a reaction, can enhance its rate. A sessile drop of reactant dispensed on the solid deforms it because of surface tension. This deformation releases excess surface free energy which can diminish the energy barrier, thereby enhancing the rate. This new mechanism may throw new light on dynamic aspects of cell-substrate interactions like cell locomotion, on action of soft robots and in design of implantable sensors.

**Introduction:**

While enhancement of rate of a chemical or biochemical reaction remains a fundamental objective in many engineering and scientific problems, parameters available for accomplishing it are not many. For a non-zero$^{th}$ order reaction, increase in concentration of a reactant (or the partial pressure for a gaseous reactant) enhances collision probability of molecules (1); increase in temperature and increases the probability of crossing over the energy barrier; increasing the interfacial area diminishes the mass transfer limitations, thereby increasing the availability of reactants at the reaction site (2); and a catalyst (or enzyme for a biochemical reaction) provides an alternate reaction path that reduces the energy barrier for a reaction to occur, thereby increasing the rate. All these effects enhance the rate of a reaction. In many situations however, these parameters may not be amenable for alteration. Here, for the first time we show that the rate of a reaction, occurring at the interface of a liquid and a solid substrate can be increased by making use of a deformable substrate. When a liquid is dispensed on a partially wettable substrate, it beads up as a sessile drop, the vertical component of surface tension of which exerts a pulling load on the substrate (3). A soft enough substrate deforms resulting in a ridge at the three-phase contact line (4-6). Surface tension driven deformation in soft solid is known as elastocapillary effect which has been studied in the context of blunting of sharp corner of a soft solid (7), flattening of the surface of a curved solid (8), estimating solid-liquid interfacial tension (9), durotaxis of liquid drop (10) and phase separation at the line of contact of two adherents (11). In biology, effect of substrate deformability has been examined in the context of cell aggregation and locomotion on compliant substrates (12), particle uptake by both epithelial cells and macrophages (13) and even in the context of substrate indentation and active mechano-transduction (14). While these examples, in general, pertain to physical effects, here we show that elasto-capillarity in soft solids can affect also the kinetics of a surface reaction. For a crosslinked, elastomeric solid, surface tension driven stresses (6,15) at the solid-air and solid-liquid interfaces lead to excess free energy which can diminish the energy barrier for a reaction. We have demonstrated this effect by carrying out reduction of aqueous solution of a metallic salt, e.g. Gold, Silver by a reducing agent, e.g. silicone hydride (SiH) present as a constituent of the curing agent, used for crosslinking the oligomers of poly(dimethylsiloxane) (PDMS).

**Results and Discussions:**

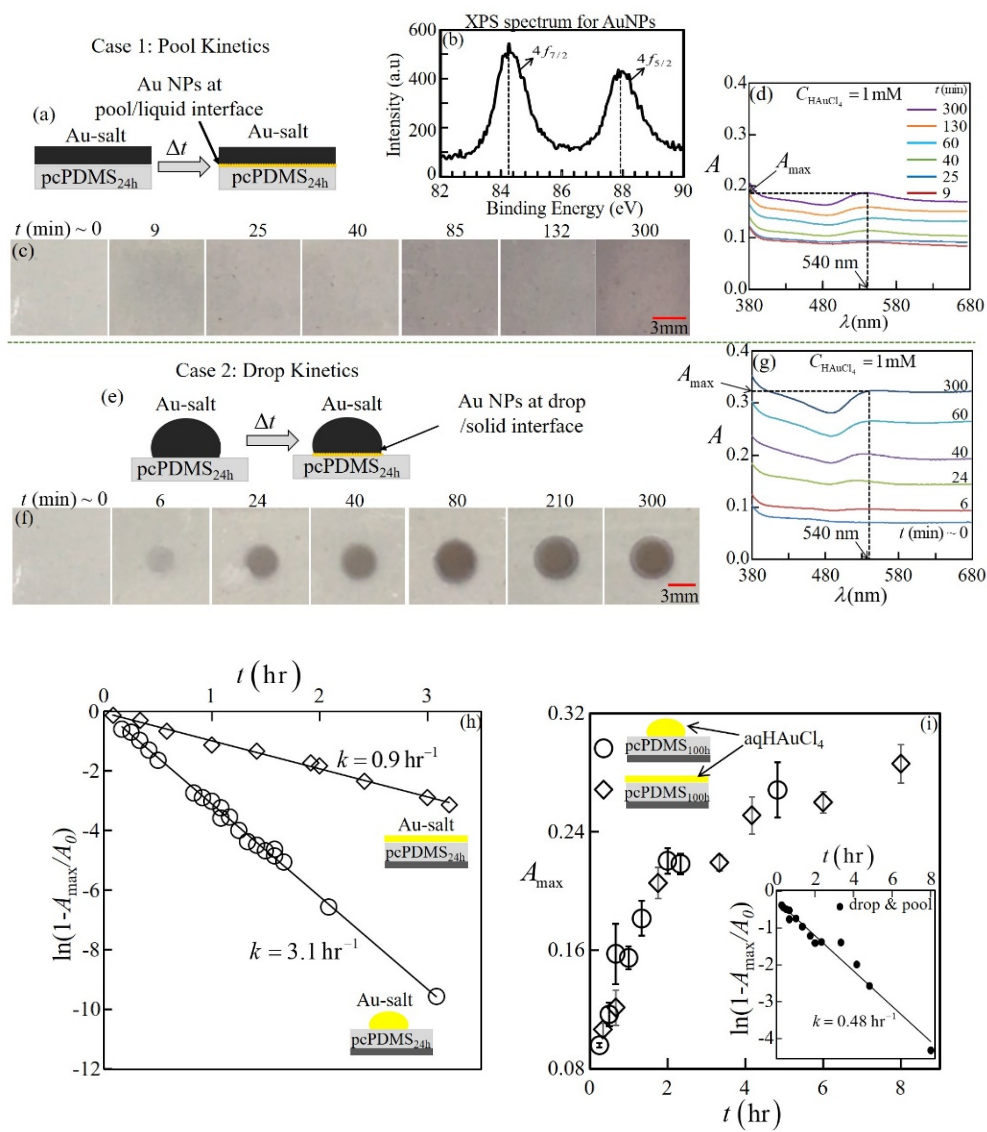

**Figure 1. Reduction of Gold salt on soft PDMS substrate. (a)** A pool of (5 ml) of the aqHAuCl₄ ($C_{HAuCl_4} = 1.0$ mM) solution was poured on the pcPDMS$_{24h}$ substrate. The salt was reduced by Silicon Hydride (SiH) present on the substrate, thereby generating Gold nanoparticles (AuNPs). **(b)** XPS spectra for these AuNPs show peaks at 87.8 eV and 84.2 eV, correspond to Binding Energies of $4f_{5/2}$ and $4f_{7/2}$ electrons of Au. **(c)** The sequence of images depicts the evolution of AuNPs at the salt-substrate interface over time. **(d)** The progress of the reaction was monitored by obtaining the UV absorbance spectra of the interface at different time steps. The maximum absorbance, $A_{max}$ occurred at 540 nm. **(e-g)** A similar set of experiments were carried out by dispensing a drop of aqHAuCl₄ ($C_{HAuCl_4} = 1.0$ mM) on the pcPDMS$_{24h}$ substrate. **(h)** $A_{max}$ vs. $t$ data for drop and pool experiments, scaled as $\ln(1 - A_{max}/A_0)$ were plotted against $t$ to yield the reaction rate constant, $k$ from the slope. **(i)** aqHAuCl₄ ($C_{aqHAuCl_4} = 3$ mM) was dispensed also on pcPDMS$_{100h}$ substrate in both drop and pool modes. The $A_{max}$ vs. $t$ data for both cases superimposed resulting in an identical rate constant.

*The Soft Substrate*

It has been shown (16,17) that when a crosslinked film of PDMS is immersed inside an aqueous solution of Chloroauric acid (HAuCl₄), SiH moieties present at the surface of the elastomer reduces the salt producing Gold nano particles (AuNPs). This reaction is slow enough that it occurs over a week (17). To examine, if softness of the substrate would enhance the reaction rate, we carried out this reaction on thin films (thickness, $h = 23 \pm 2$ μm) of partially crosslinked poly(dimethylsiloxane) (pcPDMS) prepared on plasma oxidized microscope glass slides (Materials and Methods). For a thin liquid film of crosslinkable PDMS, i.e. oligomer mixed with the crosslinking agent (PDMS$_x$), the rate of crosslinking depends on the surface chemistry of the substrate (18) on which the film is formed. The rate decreases for a substrate surface rich with SiO$_2$ e.g. that of a plasma oxidized glass. In fact, at 25°C temperature, whereas a thick film of PDMS (>100 μm) crosslinked almost completely within 24 hours, a film of thickness $h \sim 23$ μm, required 7-10 days for complete crosslinking. Thus, in our experiment, curing a thin film at 25°C allowed preparation of sufficiently soft PDMS films without decreasing the quantity of curing agent that contained the SiH. Softness of the film could be varied by crosslinking it for different duration: 24 to 100 hours, within which limited amount of SiH was consumed leaving excess of it available for any subsequent reaction. pcPDMS films crosllinked <<24 hours at 25°C were flowable; when a drop of HAuCl$_4$ was dispensed on it, the PDMS engulfed it forming a thin coating over the drop (fig. S1). This phenomenon was not of interest. For films crosslinked between 24 to 72 hours, the effect of elastocapillarity was prominent, i.e. when a sessile drop of aqHAuCL$_4$ was dispensed on it, the film deformed at the three-phase contact line forming a ridge. For pcPDMS films cured >72 hours, effect of elastocapillarity diminished as these films were less deformable.

Since films as thin as $h = 23 \pm 2$ μm could not be used for rheology experiments, the modulus of these films was estimated indirectly by obtaining wavelength of elastic instability that appears when the film is critically confined and finger like patterns appear at the interface. For soft enough films, because of elastocapillarity, wavelength of the instability patterns varies with both surface tension and modulus of the films (19,20). From this analysis, the Young's modulus of pcPDMS films crosslinked for 24h, 48h and 72h were estimated as $E = 1.62 \pm 0.03$, $3.2 \pm 0.04$ and $7.5 \pm 0.02$ kPa respectively (fig S2). Noting that the surface tension, $\gamma_{HAuCl_4}$ of aqueous solutions of Chloroauric acid, (aqHAuCl4) of different molarity, $C_{HAuCl_4} = 0.1 - 54.0$ mM was $\gamma_{HAuCl_4} = 71.7 \pm 0.3$ mJ/m² (fig S3), the elastocapillary length of a pcPDMS$_{24h}$ was estimated as $\gamma_{HAuCl_4}/E_{24h} \sim 44$ μm. Thus, the elastocapillaity induced deformation of pcPDMS$_{24h-72h}$ films was expected to be optically visible, measurable (15) and significantly larger than molecular lengthscale.

*Kinetics of reduction reaction of Gold and Silver salt*

To examine kinetics of reaction occurring on the surface of these films two different experiments were done. In the first case (fig 1a-c), the film surface was flooded with a pool of (~5ml) aqHAuCl4 ($C_{HAuCl_4} = 1.0 - 54$mM). Gold nano-particles (AuNPs) appeared at the interface via the reaction (16,21) between Si-H and the HAuCl4:

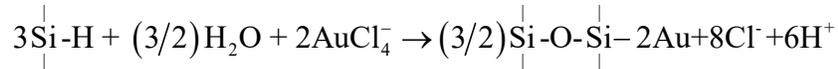

$$3\overset{|}{\underset{|}{Si}}\text{-H} + (3/2)H_2O + 2AuCl_4^- \rightarrow (3/2)\overset{|}{\underset{|}{Si}}\text{-O-}\overset{|}{\underset{|}{Si}}\text{–} 2Au + 8Cl^- + 6H^+$$

which resulted in evolution of pink color at the interface. X-ray photoelectron spectroscopy of the surface showed appearance of peaks at 87.8 eV and 84.2 eV, correspond to Binding Energies of Au 4f$_{5/2}$ and Au 4f$_{7/2}$ electrons (22) (fig 1b and fig S4). Evolution of AuNPs with time was examined by in-situ measurement of absorbance spectra (supplementary text S1) inside a UV-vis microwell plate reader (fig 1d). In the second case, the reaction was carried out between sessile drops of aqHAuCl4 and the SiH present in PDMS substrate (fig 1e-g). The sequence of images shows the evolution of the interface of the drop and the PDMS$_x$. To obtain the absorbance of the interface, the PDMS film with the drop of aqHAuCl4 on it, was placed inside the microwell plate such that the drop was placed right at the center of each well. In the absorption spectra, the peak absorbance ($A_{max}$) occurring at 540 nm was noted as a function of time, $t$, which yielded a measure of the reaction rate.

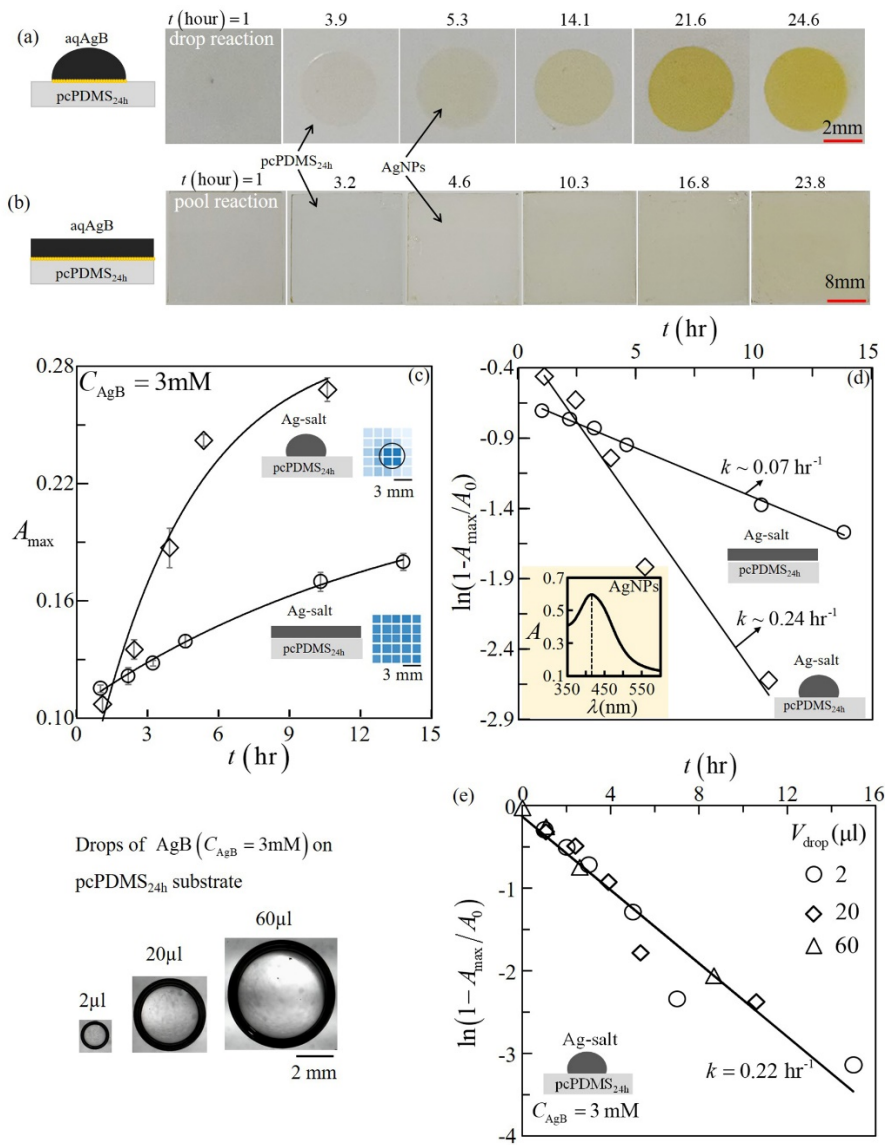

**Figure 2. Reduction of Silver salt on soft PDMS substrate. (a) Sequence of optical images shows evolution of silver nanoparticles (AgNPs) at the interface of a sessile drop (20 μl) of aqAgB ($C_{aqAgB} = 3\,\text{mM}$) and pcPDMS$_{24h}$ substrate following reaction between the salt and SiH present in the substrate. (b) The images represent evolution of AgNPs for a pool (5 ml) of aqAgB dispensed on the pcPDMS$_{24h}$. (c) At every time step, the UV absorbance of this**

**substrate was obtained in batch mode (supplementary text S1) and the maximum absorbance, $A_{max}$ (inset of plots d) was plotted a function of time $t$ for both drop and pool experiments. (d) The $A_{max}$ data were then scaled as $\ln(1 - A_{max}/A_0)$ and plotted against $t$ to yield the reaction rate constant, $k$ from the slope. (e) Drops of three different volumes: 2, 20 and 60 µl were dispensed on the pcPDMS$_{24h}$ substrate and $A_{max}$ vs. $t$ data were obtained. The plot of $\ln(1 - A_{max}/A_0)$ vs. $t$ shows that data from all three drops fall on a master line.**

For different initial salt concentration, $C_{HAuCl_4} = 0.5 - 3.0$ mM, $A_{max}$ data showed that the reaction proceeded to different extent with time. So, the reaction rate was dependent on $C_{HAuCl_4}$ (fig S5). The reaction between HAuCl$_4$ and SiH actually happens in two different steps (22): diffusion of $AuCl_4^-$ into the PDMS network at the vicinity of the interface and then its reaction with the SiH present at the neighborhood. For the pcPDMS surface, expecting SiH to be present in excess and not limited by diffusion, the reaction rate was considered to be independent of SiH concentration, $[SiH]$. Preliminary analysis showed that the reaction was first order with respect to HAuCl$_4$ concentration, $C_{HAuCl_4} = C_{HAuCl_4}|_{t=0} - [Au]$; furthermore, considering $A_{max}$ varies linearly with concentration of AuNPs, $[Au]$, $A_{max}$ was expected to vary with time $t$ as $A_{max} = A_0(1 - \exp(-kt))$. The reaction rate constant, $k$ was estimated by fitting $\ln(1 - A_{max}/A_0)$ against $t$ with $A_0$ as a fitting parameter (fig 1h, fig S5-6). For the pool and drop reaction with $C_{HAuCl_4} = 1$ mM, the reaction rate constant was obtained as $0.9 \pm 0.09$ hour$^{-1}$ and $3.2 \pm 0.6$ hour$^{-1}$ respectively. The standard deviation in both sets of data correspond to three different sets of experiments performed in identical conditions. These observations were further confirmed by carrying out experiments with other concentrations of aqHAuCl4: $C_{HAuCl_4}$ = 0.5 and 3.0 mM, for which too $k_{drop}$ was higher than $k_{pool}$ (Table 1). It is worth noting though, that with variation in $C_{HAuCl_4}$, $k$, for neither mode of reaction, was found to remain invariant, as was expected for a 1$^{st}$ order reaction. Nevertheless, dependence of $k$ on $C_{HAuCl_4}$ for both these cases was rather weak, $k_{pool/drop} \sim C_{HAuCl_4}^{0.2-0.33}$, signifying that in this range of values of $C_{HAuCl_4}$, the order of the reaction was slightly higher than 1. On average, the rate constant for drop reaction was @3.24 times that of the pool reaction. When the same reaction was carried out on a 23 μm PDMS film crosslinked for 32 -60 hour, this enhancement decreased (fig S7) and for the one crosslinked for 100 hour, pcPDMS$_{100h}$, this enhancement was missing as $k$ for both drop and pool experiment was found 0.48 hour$^{-1}$ (fig 1i). To examine generality of this observation, aqueous solution of Silver Benzoate (aqAgB) of different concentration, $C_{AgB}$ = 3.0 mM was dispensed on pcPDMS$_{24h}$ surface in both as pool (5 ml) and drops (20 μl) (fig 2a,b and fig S8). Here too, Ag nanoparticles appeared and from the absorbance data (obtained by batch measurement, section S1), the rate constants were obtained as, $k_{pool}$ = 0.07 hour$^{-1}$ and $k_{drop}$ = 0.24 hour$^{-1}$ respectively (fig 2c,d). Thus, enhancement of the reaction rate in the drop mode was ~3.4 times over the pool reaction. Increase in temperature could not account for this enhancement, because that would require the drop temperature to be increased by ~33° C compared to that of the pool (supplementary text S2), whereas the reaction was maintained at room temperature. To examine if drop volume would affect the rate, 2 - 60 μl of aqAgB drops were dispensed on pcPDMS$_{24h}$ surface, so that their initial base diameter on the substrate increased from 1.5 to 5.0 mm. Here $k$ was same: $0.22 \pm 0.02$ hour$^{-1}$ (fig 2e) for all cases. Thus, neither the drop volume, nor the area of contact between liquid and solid affected the extent of enhancement of the reaction rate in the drop vis-à-vis the pool mode. This observation suggested that for this range of liquid volume, used in experiments, the interface between the liquid and the pcPDMS$_{24h}$ substrate, was chemically and energetically uniform. In fact, evolution of the specific colours in the sequence of images in figures 1(f) and 2(a), corresponding to evolution of AuNPs and AgNPs, both show that the respective reactions occurred uniformly through-out the interface, thus confirming the above point.

| $C_{HAuCl_4}$ (mM) | 0.5 | 1.0 | 3.0 |
|---|---|---|---|
| $k_{pool}^{24h}$ (hour$^{-1}$) | $0.73 \pm 0.05$ | $0.9 \pm 0.1$ | $1.4 \pm 0.25$ |
| $k_{drop}^{24h}$ (hour$^{-1}$) | $1.9 \pm 0.2$ | $3.2 \pm 0.6$ | $3.02 \pm 0.05$ |

**Table 1. The data show reaction rate constants as obtained from experiments with aqHAuCl$_4$ of different concentration on pcPDMS$_{24h}$ surface.**

In order to examine if the enhancement is rate is due to any contamination of the liquid dispensed on the pcPDMS$_{24h}$ (thickness ~23 μm) surface over time, because of diffusion of either a species from the substrate or that of Gold nano particles (AuNPs) that emerge due to the reaction, three sets of experiments were carried out. First, several drops of DI water and that of aqHAuCl$_4$ ( $C_{HAuCl_4} = 1.0$ mM) were placed on this surface and was allowed to react with it for different duration: 5-120 minutes. From these drops, using a micropipette, small quantity (~5 μl) of the liquid was withdrawn and placed on a glass surface coated with 1H1H2H2Hperfluorooctyltrichlorosilane (FC) molecules. In figure 3 sequence of images (a) and (b) represent respectively water and aqHAuCl$_4$ drops on the FC surface. The images show that contact angle of these drops varied insignificantly: for water it decreases from 112º to 108º and for

aqHAuCl4 it decreases from 111° to 105° respectively. These results suggest that the drops did not get contaminated any significantly by the substrate surface. To probe this aspect further, in the second set of experiments, surface tension, $\gamma$ of aqHauCl4 liquid, at its initial state and that reacted for 120 min were subjected to hanging drop method (figure 3c), which showed that $\gamma$ changed insignificantly from 71.8 mN/m to 72.2 mN/m. In the third set of experiment, samples of aqHauCl4 ($C_{HAuCl_4} = 3.0$ mM) liquid before and after the reaction was subjected to UV-vis absorbance spectroscopy which showed that the absorbance data for both liquids superimposed (figure 3d). Absence of a peak between 525 and 550 nm signified absence of AuNPs in the salt solutions. Thus, these results conclusively showed that the substrate surface didn't contaminate the aqHAuCl4 drop dispensed on pcPDMS surfaces, nor there could any Marangoni flow be set in the drop that could account for the enhancement in reaction rate in the drop mode of reaction with respect to that for the pool.

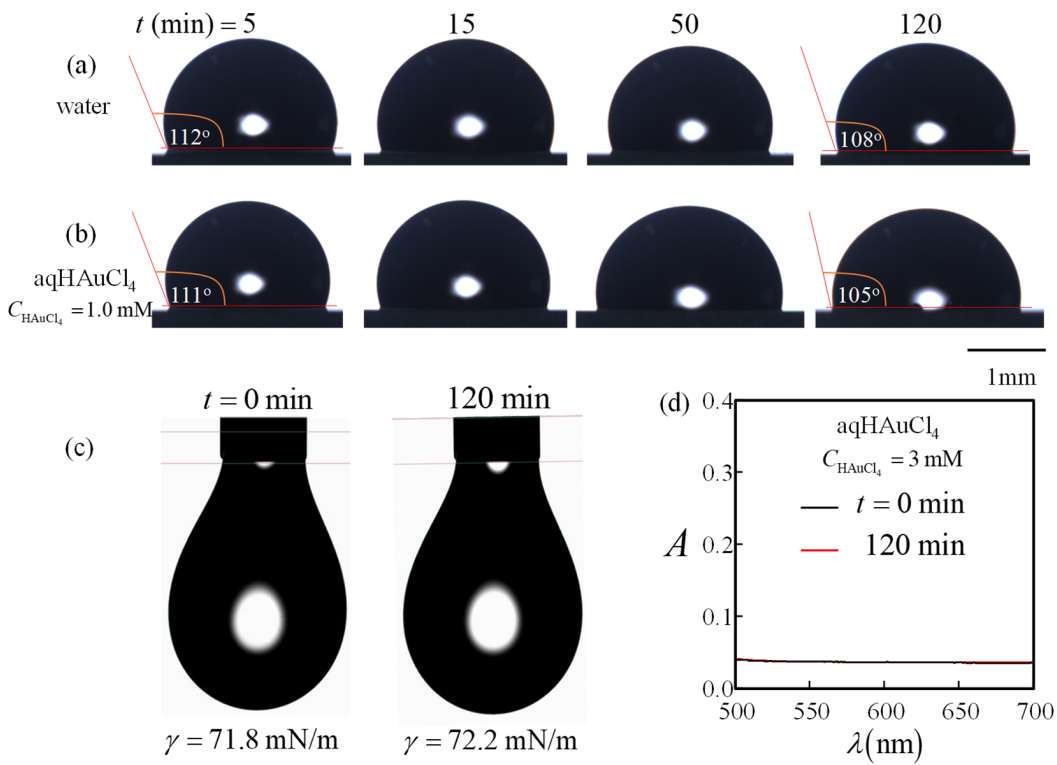

**Figure 3.** (a, b) Drops of water (a) and aqHAuCl$_4$ ($C_{HAuCl_4} = 1$ mM) (b) were kept in contact with pcPDMS$_{24h\_23\mu m}$ surface for 5, 15, 50, and 120 minutes, following which ~ 5 µl of liquid from the respective drops was withdrawn and placed on a glass surface coated with FC molecules. (c) Surface tension of aqHAuCl$_4$ ($C_{HAuCl_4} = 1$ mM) solution at initial state, i.e. $t = 0$ min and after reaction through 120 min, was estimated using hanging drop method and was found to be 71.8 mJ/m² and 72.2 mJ/m² respectively. (d) The above two liquid samples were subjected to UV-Vis spectrophotometry, which showed revealed the absence of an absorbance peak between 520 and 550 nm, indicating absence of any gold nano-particles (AuNPs) in the liquid.

*Estimation of Surface Stresses*

In fact, the reaction conditions for the drop and pool modes were essentially identical, except that the sessile drop was accompanied by a three-phase contact line, absent in the pool mode and therefore the surface profile of the PDMS film at the periphery of the drop, particularly the elastocapillarity induced deformation of the film was examined. The side view of water and aqHAuCl$_4$ drops on pcPDMS$_{24h}$ surface (fig 4a-b and corresponding magnified images) suggests occurrence of deformation of the soft substrate along the periphery of the contact area. Such deformation was absent for a sessile drop of aqHAuCl$_4$ on a completely crosslinked (ccPDMS) surface (fig 4c). For the pcPDMS$_{24h}$ film this deformation was large enough that the Young's equation involving macroscopic contact angle could not describe the balance of forces at the three-phase contact line. It has been shown that at the limit $\gamma_{HAuCl_4}/E \gg a$ ($a$: molecular lengthscale), for small enough bulk stresses, and in the absence of any long range interaction between interfaces (6), the tangents to the respective interfaces, represented by the surface tension $\gamma_l$ and surface stresses: $\Upsilon_s$, $\Upsilon_{sl}$ form the Neumann's triangle (23). Neumann's triangle corresponds to minimum free energy of the solid-liquid system that leads to balance of forces at the vicinity of the three-phase contact line (6,15,24). Here $\Upsilon_s$ and $\Upsilon_{sl}$ are the surface stresses associated with solid-air and solid-liquid interface respectively. In figure 4d the portion of the profile with positive curvature belongs to the liquid drop while that with the negative curvature delineates the deformed substrate. The point of inflection defines the contact of the two circles that nearly fit these two portions of the profile and it corresponds to the three-phase contact line (the circles were drawn for the purpose of drawing these tangents only). Therefore, the tangents $\gamma_l$ and $\Upsilon_s$ were drawn at this point along the liquid-air and solid-air interfaces respectively. To draw the third tangent, i.e. $\Upsilon_{sl}$, the symmetry of the ridge was examined by two different methods. The liquid was gently removed by blowing dry Nitrogen gas and the pcPDMS substrate was completely crosslinked by heating at 120°C for 2 hours. A thin rectangular strip of the film across the circular foot-print of the drop (rest of the film was gently scrapped out with a razor blade) was examined under optical microscope. Two tangents, drawn at the vicinity of the ridge vertex (fig 4e) formed similar angle $\sim 54.2^o \pm 0.5^o$ with the vertical line passing through it; This observation suggested that the ridge was symmetric. In the second method, the crosslinked film at the vicinity of the ridge was examined under optical profilometer (fig 4f); tangents to this profile too formed $\sim 52^o \pm 1^o$ with the vertical line at both sides, thereby showing the symmetric nature of the deformation (4,5). Both fig 3e and 3f indicate that the bulk stresses in the ridge were relaxed. Therefore, in figure 3d, at the liquid-pcPDMS interface, tangent $\Upsilon_{sl}$ was drawn symmetric to $\Upsilon_s$. The horizontal and vertical components of these surface stress values were then balanced to obtain the $\Upsilon_s$ and $\Upsilon_{sl}$ from the known value of $\gamma_l$.

|  | water on pcPDMS$_{24h}$ | aqHAuCl$_4$ on pcPDMS$_{24h}$ |
|---|---|---|
|  | $t = 2$ hour | $t = 2$ hour |
| $\gamma_l$ (mN/m) | 72.8 | 71.7 |
| $\Upsilon_s$ (mN/m) | 3.9 | 71.6±0.1 |
| $\Upsilon_{sl}$ (mN/m) | 69.3 | 2.9±0.1 |
| $W$ (mJ/m$^2$) | 7.36 | ~140 |

**Table 2. The surface stresses $\Upsilon_s$, $\Upsilon_{sl}$ and the excess free energy of the interface, $W$ was estimated by analysing the profile of the ridge as in figure 3.**

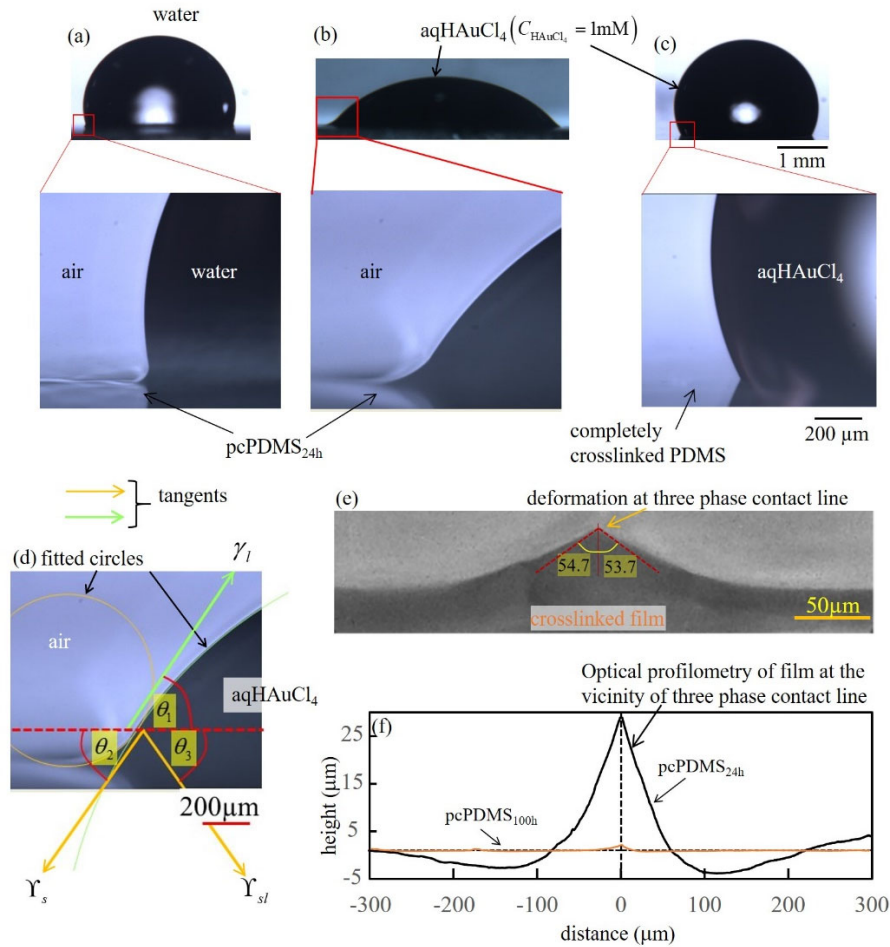

**Figure 4. Estimation of elasto-capillary effect induced excess free energy. (a-b)** The optical images show side views of drops of water and aqHAuCl$_4$, placed on pcPDMS$_{24h}$ films; the images were captured after ~3 hours of contact of the drop with the film. **(c)** The images shows a similar drop of aqHAuCl$_4$, placed on a ccPDMS film. The magnified views capture the deformation of the films at the vicinity of the three-phase contact line. **(d)** Side view of image (b) was analyzed by drawing tangents at interfaces: liquid-air, solid-liquid and solid-air; the tangents define the corresponding interfacial tension ($\gamma_l$) and interfacial stresses ($\Upsilon_s$ and $\Upsilon_{sl}$). **(e)** The drop was gently removed and the solid substrate was completely crosslinked by heating at 100 °C for 2 hours. The ridge, present along the drop periphery,

was captured by optical microscopy. (f) The contour of the ridge was captured also by optical profilometry.

*Estimation of Excess Free Energy*

For the controlled case of drop of water / aqHAuCl4 on a ccPDMS surface, in the absence of elastocapillarity induced deformation, the interfacial tension was obtained by applying Young's equation: $\gamma_{sl} = \gamma_s - \gamma_l \cos\theta$. Noting that surface tension of water and aqHAuCl4 are $\gamma_{DI} = 72.7$ mN/m and $\gamma_{HAuCl_4} = 71.7$ mN/m respectively, that of ccPDMS, $\gamma_s = 22$ mN/m, and $\theta_{DI} = \theta_{HAuCl_4} = 109° \pm 1°$ (fig S9), the solid-liquid interfacial tension, $\gamma_{sl}$ for DI water and HAuCl4 were calculated as 45.8 mN/m and 44.6 mN/m respectively. The excess free energy of the interface, $W = \gamma_s + \gamma_l - \gamma_{sl}$ was then estimated for DI water-ccPDMS and aqHAuCl4-ccPDMS, both as ~ 49 mJ/m². For the drop of DI water dispensed on pcPDMS$_{24h}$ film, as in figure 4a, the macroscopic contact angle still remained $\theta_{DI} \sim 110°$, however presence of elastocapillary effect altered the surface stresses at the vicinity of the three-phase contact line. Analysis of the side images of water as in figure 3a yielded $\Upsilon_s$ and $\Upsilon_{sl}$ as 3.9 mN/m and 69.3 mN/m respectively. For the drop of aqHAuCl4 ($C_{HAuCl_4} = 1.0$ mM, $\gamma_l = 71.7$ mN/m) in contact with the pcPDMS$_{24h}$ film for 2 hours (during which the macroscopic contact angle decreased from $\theta_{aqHAuCl_4} \sim 111° \pm 3°$ to ~ 66°), $\Upsilon_s$ and $\Upsilon_{sl}$ were estimated as ~71.6 mN/m and 2.9 mN/m respectively (25). For the two systems, involving two different liquids, $\Upsilon_{sl}$ values were expected to be different. However, $\Upsilon_s$ too were different, at the vicinity of the respective drops, as the respective solid surfaces were at different energy (differently stressed) and entropic state with different Helmholtz free energy

(26) (supplementary text S3). Furthermore, in contrast to the water drop, for aqHAuCl4, some NPs diffused from the S-L interface to the S-A interface (fig S11). In essence, the pcPDMS$_{24h}$-air interface at the vicinity of drops of two different liquids were essentially two different interfaces. Consequently, their surface tension values too were expected to be different.

Surface tension and surface stress of a solid are related by the following equation (27): $\gamma_s = \Upsilon_s - A\frac{d\gamma_s}{dA}$, $\gamma_{sl} = \Upsilon_{sl} - A\frac{d\gamma_{sl}}{dA}$, in which the second term represents the change in free energy because of reversible elastic strain in the surface area of the interface and $A$ is the surface area. For a crystalline solid, atoms are constrained to lattice nodes; therefore, when the surface of the crystal is subjected to a strain, the chemical potential of atoms at the surface alters from those at the bulk. In other word, the free energy of the surface varies with the area strain and the surface tension differs from the surface stress (28). In the other limit for a liquid, chemical potential of molecules at the bulk remains equal to those at the surface irrespective of any surface strain. So, for liquid $\gamma_l = \Upsilon_l$. For an amorphous solid like a crosslinked elastomer, at temperature >> glass transition temperature, $T_g$ ($T_g|_{PDMS} = -140°\,C$) segments of the polymer chain between crosslinking points remain mobile like a liquid (29,30); mobility increases with decrease in the crosslinking density. Therefore, chemical potential of segments at the bulk equilibrate to those at the surface and the surface free energy remains unaltered with strain: $A\frac{d\gamma_s}{dA} = 0 = A\frac{d\gamma_{sl}}{dA}$. Then the solid-air and solid-liquid interfacial tension equate to the corresponding surface stresses: $\gamma_s \approx \Upsilon_s$ and $\gamma_{sl} = \Upsilon_{sl}$. For aqHAuCl4-pcPDMS$_{24h}$ system, $\gamma_s$ and $\gamma_{sl}$ were then equal to 71.6 mN/m and 2.9 mN/m respectively and the corresponding excess free energy was $W_{HAuCl_4}^{drop} \sim \gamma_s + \gamma_l - \gamma_{sl} = 140.4$ mJ/m². In contrast, for a pool of aqHAuCl4 dispensed on the pcPDMS24h film, in absence of a three-phase contact line, the surface stresses did not develop. Then $\gamma_s$ was same as that of unstressed PDMS surface: $\gamma_s = 22$ mN/m. Furthermore, considering that solid-liquid interfacial tension was same irrespective of the pool and drop contact, $\gamma_{sl}$ was equal to 2.9 mN/m. The excess free energy of the interface in the pool geometry was thus obtained as, $W_{HAuCl_4}^{pool} \sim 91$ mJ/m².

*Effective Activation Energy Barrier*

To examine the effect of excess free energy of interface on reaction rate, the Arrhenius expression of rate constant was considered: $k = k_0 \exp\left(-\frac{E_A}{RT}\right)$, where $E_A$ (J/mole) is the activation energy barrier. It includes the intrinsic energy barrier, $E_A^*$ and the excess free energy of the interface calculated per mole of the salt solution, $\overline{W}$. For aqHAuCl4-pcPDMS system, the excess free energy in the pool and drop geometry were estimated as (supplementary text S4) $\overline{W}_{HAuCl_4}^{pool} = 5.28$ kJ/mole and $\overline{W}_{HAuCl_4}^{drop} = 8.16$ kJ/mole respectively. Writing $k \sim \exp\left(-\frac{E_A^* - \overline{W}}{RT}\right)$, the ratio of rate constants for the drop and pool reaction at $T = 298$ K was then estimated as,

$$\frac{k_{pcPDMS}^{drop}}{k_{pcPDMS}^{pool}} = \frac{\exp\left(\frac{\overline{W}_{pcPDMS}^{drop}}{RT}\right)}{\exp\left(\frac{\overline{W}_{pcPDMS}^{pool}}{RT}\right)} = 3.2$$

This value corroborates well with the ratio 3.24 of rate constants obtained from the experiments.

To ascertain that elastocapillarity induced surface stress was essential for the enhancement in reaction rate, the pool and drop reactions were carried out on PDMS substrate in which SiH was expected to be present in excess, yet the effect of elastocapillarity was significantly diminished. A 23 μm film of PDMS crosslinked for 100 hour, i.e. pcPDMS$_{100h}$ ($E = 13.5 \pm 3.5$ kPa) was found suitable for this purpose; compared to pcPDMS$_{24h}$, a sessile drop of aqHAuCl$_4$ ($C_{HAuCl_4} = 1$ mM) deformed this film to an insignificant extent (fig 4f). The $A_{max}$ data obtained as a function of time from both the modes of reaction on this surface nearly superimposed (fig 1i) buttressing the point that in absence of elastocapillarity, the reaction rate was unaffected by the surface and in both modes the rate constant was estimated to be $k_{pcPDMS_{100h}} = 0.48$ hour$^{-1}$. At the other limit, maximum enhancement in rate was expected when $\overline{W}^{drop}$ maximizes, i.e. when $\gamma_{sl} \to 0$ mN/m. For the aqHAuCl$_4$-PDMS system, this value was 3.3 as observed in the experiments.

## *Summary*

It has been shown recently via quantum mechanical calculations that intrinsic stiffness of a crystalline material like SiC is expected to increase the rate of a surface reaction like scission of Si-C bonds under attack by H of water (31). Here, alteration in bulk stoichiometry results in release of mechanical energy that is thought to diminish the energy barrier for reaction. In contrast, we for the first time have shown here that in soft solids, elastocapillary effect enhances the rate of a surface reaction via surface tension driven deformation at its surface. In essence, this effect is akin to similar other phenomena described earlier, e.g. excess surface energy associated with drop coalescence leads to rapid motion of resultant drop on a surface (32); excess energy of wetting of the curved surface of a microchannel embedded in a soft elastic layer leads to its bulging deformation (9) and so on. Apart from reduction of metallic salt as presented here, it may be relevant for many other reactions particularly ones happening at the surface. For example, biological cells like fibroblasts migrate from softer to a stiffer surface (durotaxis). While such migration is mediated by polymerization-depolymerization of intracellular fibers like actin and actomyosin (33), softness of the substrate, because of the effect presented here is expected to enhance the actin turnover rate, leading to increased speed of the cell; the speed is expected to diminish on a harder surface in absence of the excess surface energy. Similarly, stress field inside a tumour, embedded within a soft tissue, like that of brain, is profoundly affected by the softness of the surrounding tissue (34). Elastocapillarity induced surface stress is expected to affect the growth rate of the tumor. Catalysts are generally made on porous yet rigid support. However soft and porous substrates, as for many enzyme mediated bio-chemical reactions, can possibly enhance the reaction rate over and above what is conventionally achieved. Our results show that elasticapillary effect can be harnessed also to enhance the rate of detection of an analyte by a sensor.

**Materials and Method**

**Materials:** Sygard 184 elastomer kit was purchased from Dow Corning. Chloroauric acid (extra pure) was purchased from Loba Chemie. Pvt. Ltd.. Silver benzoate and Trichloro(1H,1H,2H,2H-perfluorooctyl)silane (FC) were obtained from Sigma Aldrich. Sulfuric acid (99%) was purchased from RANKEM and hydrogen peroxide (30%) was purchased from Qualigens, Thermo Fisher Scientific. Needle of diameter 450 to 550 was purchased from local market. Milli Q (DI) water was used in all experiment.

**Method of preparation of precursor salt solution:** Aqueous solutions of Chloroauric acid, (aqHAuCl4) of desired molarity, $C_{HAuCl_4} = 0.1 - 54.0$ mM were prepared by mixing the salt in desired quantity in DI water. The surface tension, $\gamma$ of the resultant solution was measured using pendant drop method in which contour of a small drop of liquid, hanging from a vertically placed needle in air, was extracted using Data Physics OCA 35 instrument. The contour was fitted to the Young Laplace equation to obtain the surface tension of liquid. For aqHAuCl4 with $C_{HAuCl_4} = 1$ mM, the $\gamma_{HAuCl_4}$ values were obtained as, $\gamma_{HAuCl_4} \sim 71.7 \pm 0.5$ mJ/m$^2$.

**Method of preparation of pcPDMS and ccPDMS films:** Thin films (thickness, $h = 5 - 25$ μm) of partially crosslinked PDMS (pcPDMS) were prepared by spin coating (at 2000-8000 rpm for 90 sec) Sylgard 184 elastomer mixed with the curing agent in 10:1 weight ratio (referred to as PDMS$_x$) on plasma oxidized microscope glass slides. For thin film of PDMS$_x$, the rate of crosslinking reaction depends on the surface chemistry of the substrate (*18*). Particularly, the rate decreases when the film is crosslinked against surfaces having SiO$_2$, e.g. the plasma oxidized glass. Our experiments showed that at 25°C temperature, whereas a thick film of PDMS$_x$ ($> 50$ μm) crosslinks almost completely within 24 hours, a film of thickness $h \sim 25$ μm, requires about 7 days for complete crosslinking. To prepare films of different softness, liquid films of PDMS$_x$ of thickness $h \sim 23$ μm were cured at 25°C for different duration: 24 to 72 hours. Within this period, the crosslinking reaction could proceed only to a limited extent thereby consuming only limited amount of SiH present in PDMS$_x$. Thus, for pcPDMS films crosslinked through 24-72 hour, SiH was available in excess quantity at its surface for reaction with the HAuCl4. Therefore, it was considered that the rate of the reaction was independent of the SiH concentration for these pcPDMS films. Films crosllinked at 25ºC for less than 24 hours, were flowable like liquid, so that, when a drop of HAuCl$_4$ was dispensed on it, the PDMS$_x$ could engulf the whole of it forming a thin coating over the drop. This phenomenon was not of interest and therefore films crosslinked longer than 24 hours were used in experiments. For films of $h \sim 23$ μm, elastocapillarity induced deformation led to a ridge at the three-phase contact line but the PDMSx did not engulf the sessile drop of aqHAuCL$_4$ on it. For pcPDMS films cured longer than 72 hours at 25ºC, elastocapillarity induced deformation diminished following decrease in their deformability and these cases were not of interest for the work presented here.

**Acknowledgments**:

NS acknowledges G. Lohiya and M. Sriram for helping with UV-vis spectroscopy, S. Tiwari for helping with Goniometer and A. Kumar for providing a PDMS film embedded with micro-channel.

**Funding:** This work was supported by Science and Engineering Research Board (SERB), Government of India in the form of grants STR/2019/000044 and CRG/2021/00128.

**Author contributions:** Conceptualization: N.S. and A.G. Methodology: N.S. and A.G. Investigation: N.S. and A.G. Validation: N.S. and A.G. Visualization: N.S. and A.G. Funding acquisition: A.G. Data curation: N.S. Writing – original draft: N.S. Writing review and editing: A.G.

**Competing interests: Authors have no competing or conflict of interest in regard to this report.** competing interests of any of the authors must be listed (all authors must also fill out the Conflict of Interest form). Where authors have no competing interests, this should also be declared.

**Data and materials availability:** All data is available in the manuscript or in the supplementary materials.

**Supplementary Materials**
Supplementary text S1 – S4
Table S1
Fig S1 – S11
References (1 – 29)

# Supporting Online Material

## Enhancement of the Rate of Surface Reactions by Elasto-capillary Effect

Nitish Singh[1] and Animangsu Ghatak[1,2,*]

[1] Department of Chemical Engineering, Indian Institute of Technology Kanpur, 208016
[2] Center for Environmental Science and Engineering, Indian Institute of Technology Kanpur, 208016 (India)
[*] Prof. A. Ghatak, Corresponding-Author, Author-Two, E-mail: aghatak@iitk.ac.in

**List of items**

**Fluid like behavior of a pcPDMS$_{10h}$ film**

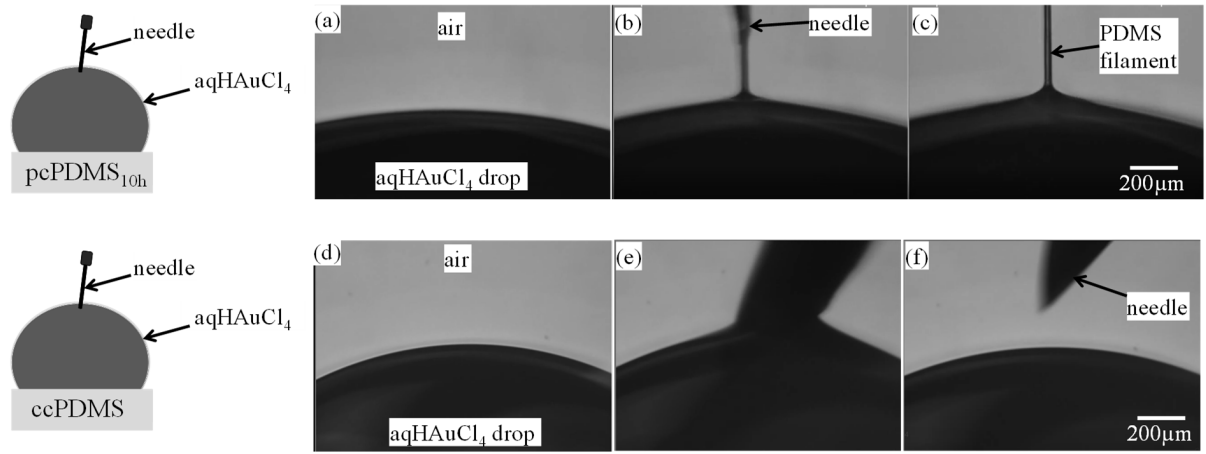

**Figure S1. (a-c)** Sequence of images shows that a thin (thickness ~23 μm) film of PDMS formed on a rigid glass slide, crosslinked for 10 hour, i.e. a pcPDMS$_{10h}$ film, actually remains fluidic. A 10μl drop of aqHAuCl$_4$ ($C_{HAuCl_4} = 1\,\text{mM}$) was dispensed on this film and was allowed to relax for 10 minutes. A syringe needle of diameter 450 μm was then inserted into the top surface of the drop and then gently withdrawn. At this, the needle did not separate smoothly from the drop but a thin liquid filament formed, illustrating that the polymer had engulfed the liquid drop. **(d-f)** A similar sequence of process was followed by dispensing the aqHAuCl$_4$ drop on a ccPDMS film. In this case, the filament did not appear and the needle separated in clean manner signifying that here the drop was not engulfed by the polymeric substrate.

**Estimation of modulus of pcPDMS$_{24h-72h}$ films:**

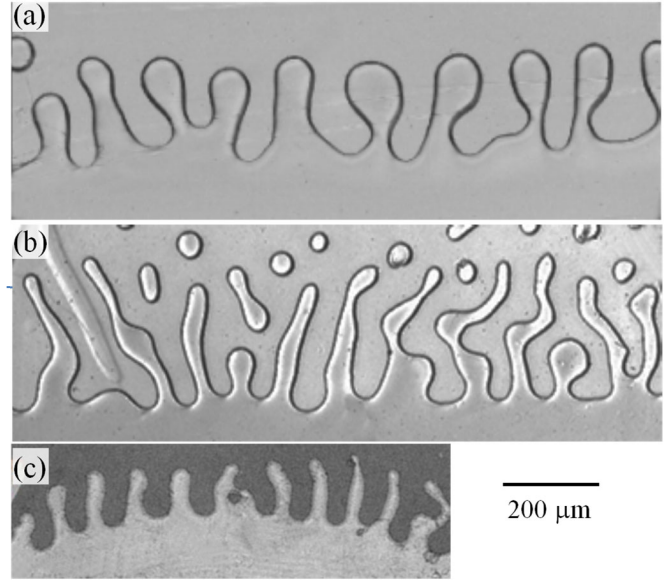

**Figure S2. (a-c) The images depict instability pattern which appear in experiments in which thin films (thickness $\sim 23\,\mu m$) of pcPDMS is brought in gentle contact with freshly cleaved mica sheet. Finger like structures appear along the line of contact between the two adherents, very much similar to confinement induced elastic instability described in references[2-5]. Images (a) to (c) correspond to pcPDMS films crosslinked for 24 hours, 48 hours and 72 hours respectively.**

Elastic modulus of pcPDMS films of thickness, $h \sim 23$ μm was estimated by obtaining wavelength of elastic instability. In essence, a thin sheet of freshly cleaved mica was gently placed on the surface of these films (strongly bonded to a rigid microscope glass slide) in the peeling geometry[2], which led to the appearance of static, uniformly separated finger like patterns along the contact line. Recognizing that these fingers were signatures of elastic instability in confined thin elastic film [3] (they were distinctly different from that in viscoelastic films in similar geometry[4], the wavelength, $\lambda$ of these patterns were obtained[4,5]. For the 23μm thick pcPDMS$_{24h}$ film, $\lambda$ was obtained as ~130μm. The shear modulus, $\mu$ of this film was then estimated by using the equation $\lambda \sim 2\pi h(\gamma/3\mu h)^{1/4}$ as deduced in references 4 and 5 relating the wavelength to the thickness, $h$ of the film, its shear modulus $\mu$, and the surface tension, $\gamma$. Putting $h = 23$ μm, $\gamma = 22\,\text{mJ/m}^2$, $\mu$ was estimated as 544 N/m². Considering these films to be incompressible, with Poisson ratio, $\nu = 0.5$, the Young's modulus was estimated as $E = 2\mu(1+\nu) = 1632$ N/m². A similar set of analysis was done for pcPDMS$_{48h}$ and pcPDMS$_{72h}$ films to obtain $E$ for the respective films as 3157 N/m², and 7483 N/m² respectively.

| Sample | $\lambda$(μm) | $h$(μm) | $\mu$(N/m²) | $E$(N/m²) |
| --- | --- | --- | --- | --- |
| pcPDMS$_{24h}$ | 130 | 23 | 544 | 1631 |
| pcPDMS$_{48h}$ | 109 | 23 | 1052 | 3157 |
| pcPDMS$_{72h}$ | 81 | 21 | 2494 | 7483 |

**Table S1. Table shows the estimated elastic modulus for 24 hour, 48 hour and 72 hour crosslinked PDMS based on wavelength measurements of adhesion-induced instability.**

**Determination of surface tension of aqHAuCl₄ solution**

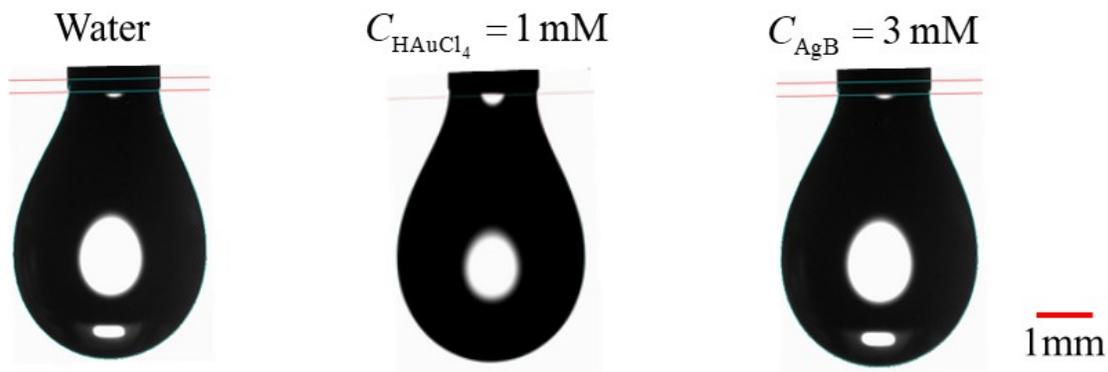

**Figures S3.** Pendant drop method was used to estimate the surface energy of water, aqHAuCl₄ ($C_{HAuCl_4} = 1.0$ mM) and aqAgB ($C_{AgB} = 3.0$ mM) respectively. A drop of the liquid was allowed to hang vertically in air from a syringe needle. Data Physics OCA 35 (goniometer) was used to extract the contour of this drop which was fitted to the Young Laplace equation. The surface energy of the three liquids were found as $\gamma_{H_2O} = 72.8$ mJ/m², $\gamma_{HAuCl_4} = 71.7 \pm 0.3$ mJ/m² and $\gamma_{AgB} = 73.4 \pm 0.1$ mJ/m² respectively.

**Section 5. SEM and TEM of Gold nano-particles (AuNPs):**

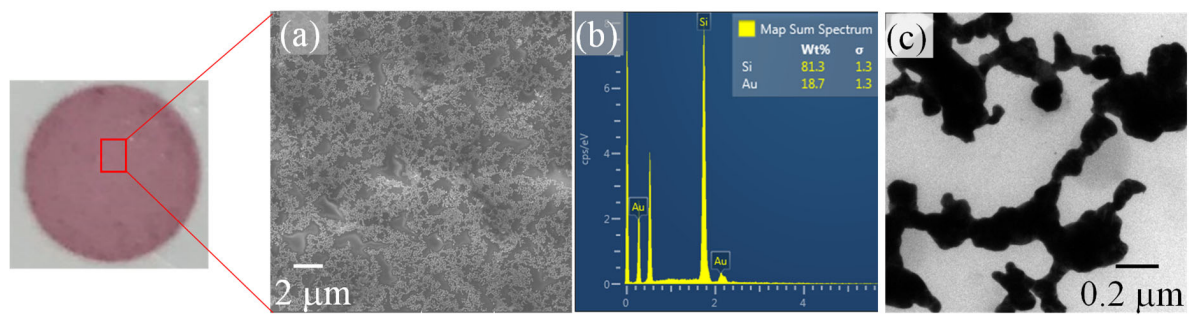

**Figure S4.** A sessile drop of aqHAuCl₄ ($C_{HAuCl_4} = 1$ mM), dispensed on a typical pcPDMS$_{24h}$ substrate was allowed to react over 24 hours, following which the liquid was gently sucked out and the surface was dried by blowing dry N$_2$ gas. The PDMS surface was then rinsed with water and ethanol and cured at 70º C for 3 hours. Evolution of purple color indicated the presence of AuNPs nanoparticles. The patch was subjected to Scanning Electron Microscopy (SEM), Energy Dispersive X-Ray Analysis (EDX) and Transmission Electron Microscopy (TEM).

**Supplementary Text S1:**

**Measurement of absorbance spectra:**

The progress of reduction reaction of the aqueous solution of the salt by SiH was monitored by obtaining the absorbance with respect to time. For this purpose, the reacted interface of the pcPDMS surface was first scanned between 380nm and 700nm to obtain the absorbance spectrum. The absorbance peak for gold nanoparticles was around 540nm, and for silver nanoparticles, it was around 410/420 nm. Absorbance as a function of time was then obtained for both drop and pool experiments by two different measurement methods: **in situ** and **batch**.

**In situ measurement of absorbance at a given wavelength:** The absorbance at the interface of the liquid drop and solid substrate was obtained as a function of time using a 24-well microwell plate. The pcPDMS substrate prepared earlier was first placed inside one of the wells of the microwell plate. A 20 µl drop of the salt solution of desired concentration was then deposited on it. To prevent the drop from evaporating, the area surrounding the drop was filled with water. This plate was inserted inside the BioTek Synergy H4 Microplate Reader to record absorbance at 540 nm at every 5-10 minutes for 3-5 hours at a single location at the interface.

Similar to the above protocol, for the pool experiment, the pcPDMS substrate was first placed inside a well of a microwell plate. 5 ml of aqueous solution of the salt was dispensed in a 6 well plate; few experiments were done using 12 well and 24 well plate for which 2.5 ml and 0.5 ml liquid was used respectively. Thus, in each case, a liquid layer of thickness ~3 mm was formed on the pcPDMS. The absorbance at 540 nm was then obtained at every 5–10 minutes for 3–5 hours. Thus, evolution of absorbance at a particular location at the interface was obtained over time.

**Batch measurement of absorbance via area scan:** In the second approach, a 20 µl drop of the aqueous solution of salt was dispensed on the pcPDMS surface. The substrate, with the sessile drop on it, was placed inside a chamber with controlled humidity to prevent any evaporation. Several such samples were prepared and for each case, the reaction between the drop and the substrate was allowed to happen for a given time. After each time step, the drop was gently removed from the substrate by blowing Nitrogen gas. The interface was then rinsed with Milli-Q water to wash off any remaining salt and was blow dried again with Nitrogen gas. The pcPDMS substrate with the reacted patch was placed inside the microwell plate. Absorbance was obtained in area scan mode, at 540 nm for Gold nanoparticles and at 410nm for Silver nanoparticles. Absorbance was obtained at 3-4 spots, each having a size of ~1.54 mm$^2$ and the average value of the absorbance was further analyzed for estimating the reaction rate constant.

A similar set of steps were followed for the pool experiment, for which absorbance was measured at 25 spots of area 1.54 mm$^2$ each.

**Maximum absorbance data with time for $C_{HAuCl_4} = 0.5 - 3.0 \, \text{mM}$**

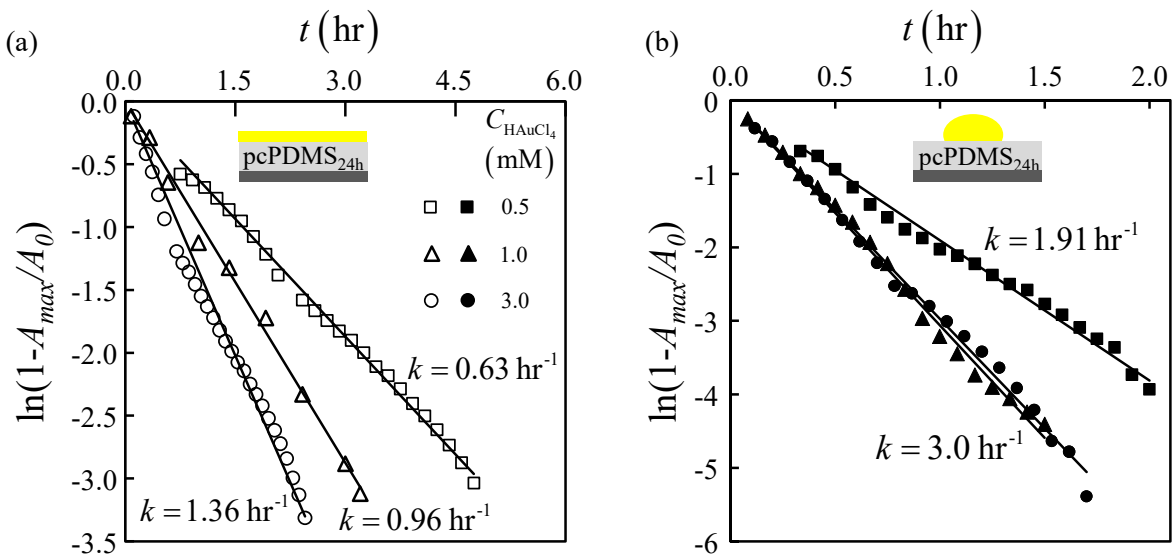

**Figure S5.** aqHAuCl₄ solutions having salt concentration varying from $C_{HAuCl_4} = 0.5 - 3.0 \, \text{mM}$ were reacted with pcPDMS$_{24h}$ substrate in both pool (a) and drop (b) mode. The maximum absorbance $A_{max}$ data scaled as $\ln(1 - A_{max}/A_0)$ were plotted against time $t$ to yield the reaction rate constant, $k$. In the pool mode, for $C_{HAuCl_4} = 0.5, 1.0$ and $3.0 \, \text{mM}$, $k$ was found to be 0.63, 0.96 and 1.36 hour⁻¹ respectively. In the drop mode, for $C_{HAuCl_4} = 0.5 \, \text{mM}$, $k$ increased to 1.91 hour⁻¹. In the drop mode, $k$ for $C_{HAuCl_4} = 1.0$ and $3.0 \, \text{mM}$ was similar ~3.0 hour⁻¹.

**Determination of reaction rate constant in repeated experiments.**

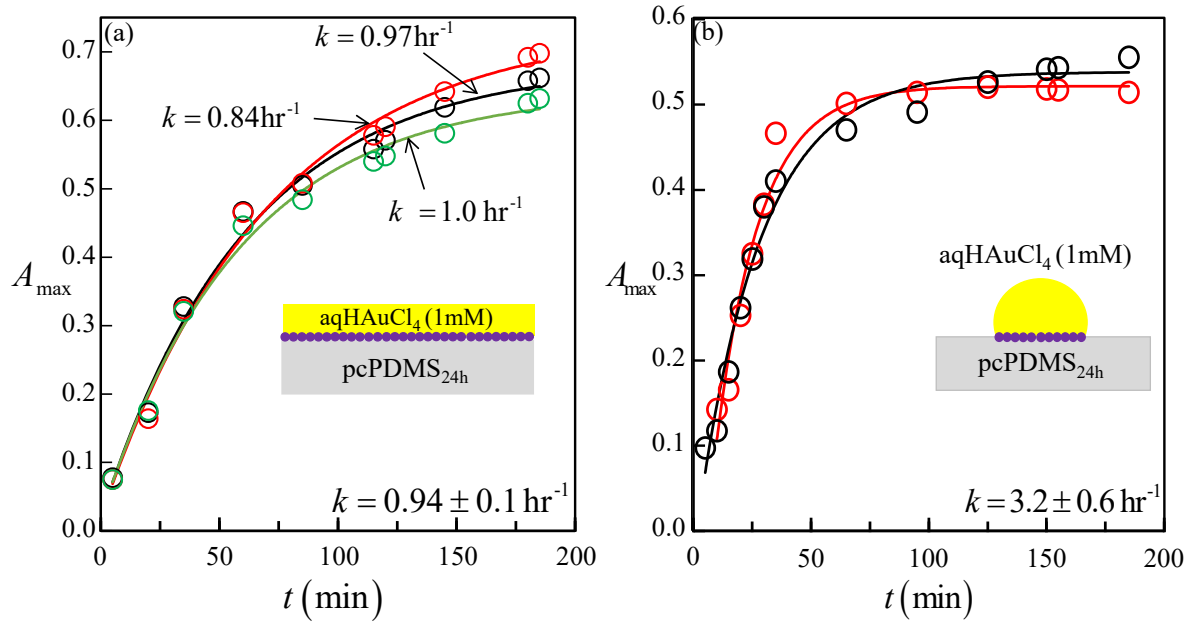

**Figure S6. (a)** The plot shows the results from experiments in which a pcPDMS$_{24h}$ film was flooded with aqHAuCl$_4$ ($C_{HAuCl_4} = 1.0$ mM) and the evolution of AuNPs were monitored 200 min in-situ via UV-vis spectrophotometry. The peak absorbance, $A_{max}$ data were then fitted to 1$^{st}$ order reaction kinetics to yield the reaction rate constant, $k$. Different colors represent three different sets of experiments which yield the rate constant as $0.94 \pm 0.1$ hr$^{-1}$. **(b)** A similar experiment was carried out by dispensing aqHAuCl$_4$ ($C_{HAuCl_4} = 1.0$ mM) as sessile drop on the pcPDMS$_{24h}$ surface. Here too, the $A_{max}$ data from two different sets of experiments were fitted to the first order kinetics, to yield the rate constant as $3.2 \pm 0.6$ hr$^{-1}$.

**Determination of reaction rate constant for substrate with different extent of crosslinking**

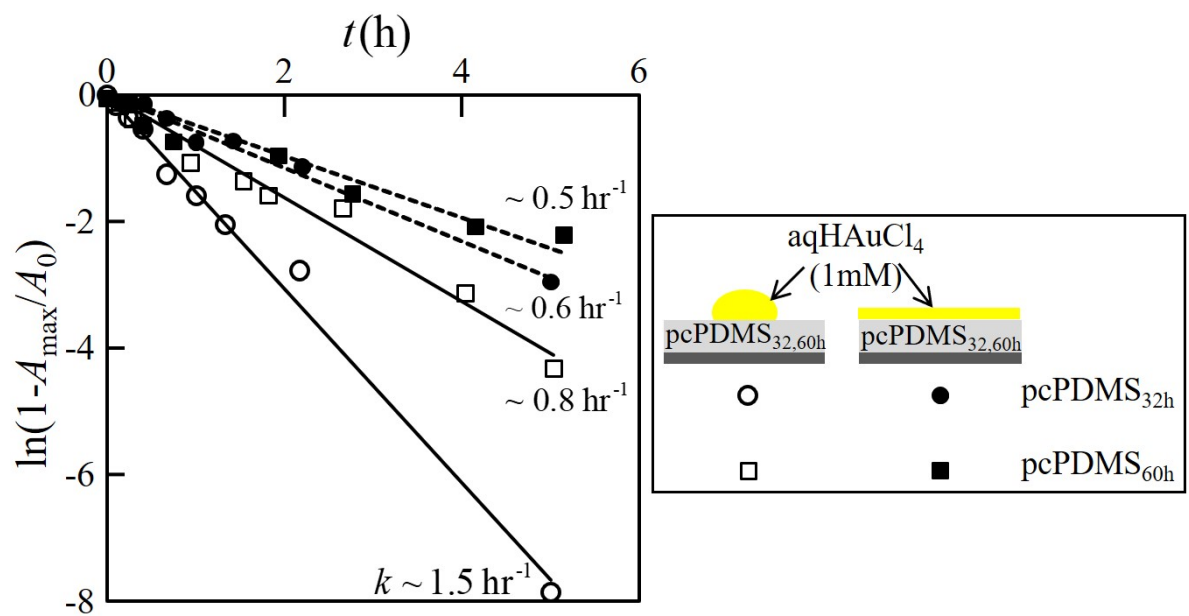

**Figure S7. (a) pcPDMS films crosslinked to 32 hours and 60 hours respectively were used as the substrate and were subjected to pool and drop experiments with aqHAuCl$_4$ ($C_{HAuCl_4}$ = 1.0 mM) as in figure 1. The peak absorbance, $A_{max}$ data were then fitted to 1$^{st}$ order reaction kinetics to yield the respective reaction rate constant, $k$. For the pool experiment, on pcPDMS$_{32h}$ and pcPDMS$_{60h}$ substrates, $k$ were estimated as 0.6 hour$^{-1}$ and 0.5 hour$^{-1}$ respectively. For the experiment in drop mode, these values were obtained as 1.5 hour$^{-1}$ and 0.8 hour$^{-1}$ respectively.**

**Kinetics of formation of silver nanoparticles (AgNPs)**

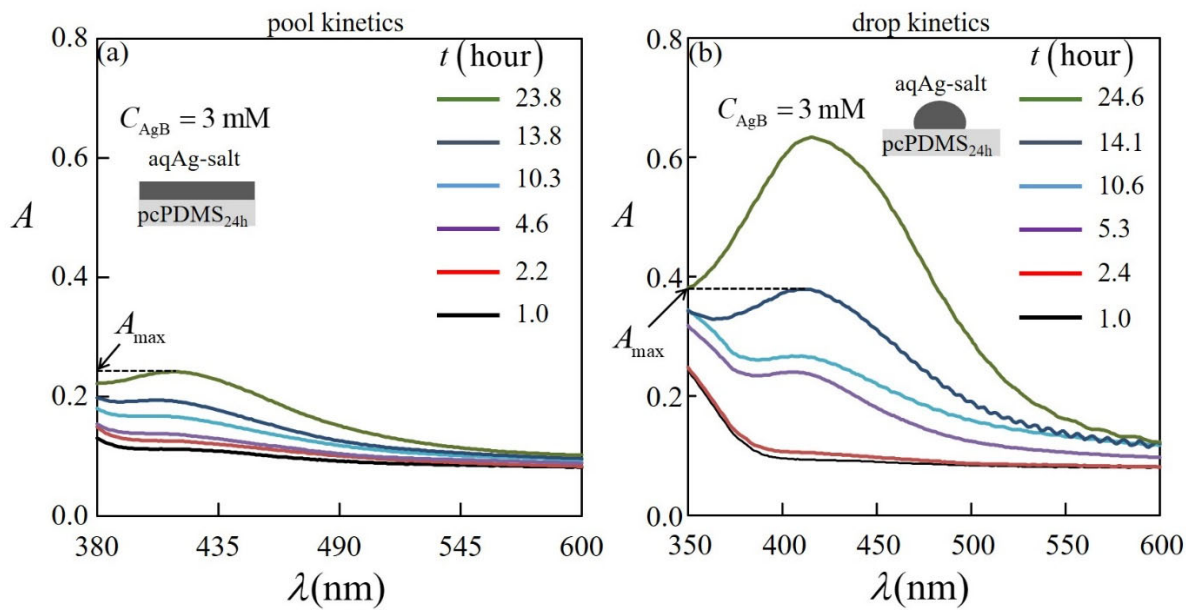

**Figure S8.** The plots show typical absorbance spectra for reaction in the pool and drop modes respectively as a function of time. (a) 20 µl drops of aqAgB ($C_{AgB} = 3\,\text{mM}$) were dispensed on 23 µm thick pcPDMS$_{24h}$ films and were allowed to react for different time durations: 1 to 24 hour. UV absorbance of the interface at each time step was obtained in batch mode. (b) pcPDMS$_{24}$ films were flooded with ~6ml of aqAgB ($C_{AgB} = 3\,\text{mM}$) liquid forming a liquid layer of thickness 3-4 mm. Here again the absorbance of the interface in the batch mode similar to the drop experiment. For the drop reaction, the peak absorbance ($A_{max}$) was found to increase more rapidly than in the pool case.

**Supplementary Text S2**

**Can increase in temperature account for the increase in the reaction rate in the drop mode over that in the pool mode?**

**The PDMS surface, on which the reaction takes place in both the drop and pool modes, remains at thermal equilibrium with the surroundings and therefore its temperature, $T$ is expected to remain constant at 25°C for both cases. It is however useful to ascertain if increase in temperature $T$ could account for the increase in the reaction rate in the drop mode over that in the pool mode, as observed in our experiments. The activation energy barrier for the reaction between HAuCl$_4$ and SiH has earlier been estimated to be 29.8 kJ/mole[6,7]. Considering that the pool reaction happens at $T = 298\,\text{K}$, calculations show that the drop reaction could occur at 3.3 times the rate as that of the pool provided the temperature would increase $T = 331\,\text{K}$ to i.e. at a temperature ~33° C higher than that of the pool. Such difference in temperature is not observed in experiments and therefore the possibility of rate enhancement via increase in reaction temperature was discarded.**

**Side views of aqHAuCl4 drops on pcPDMS surfaces:**

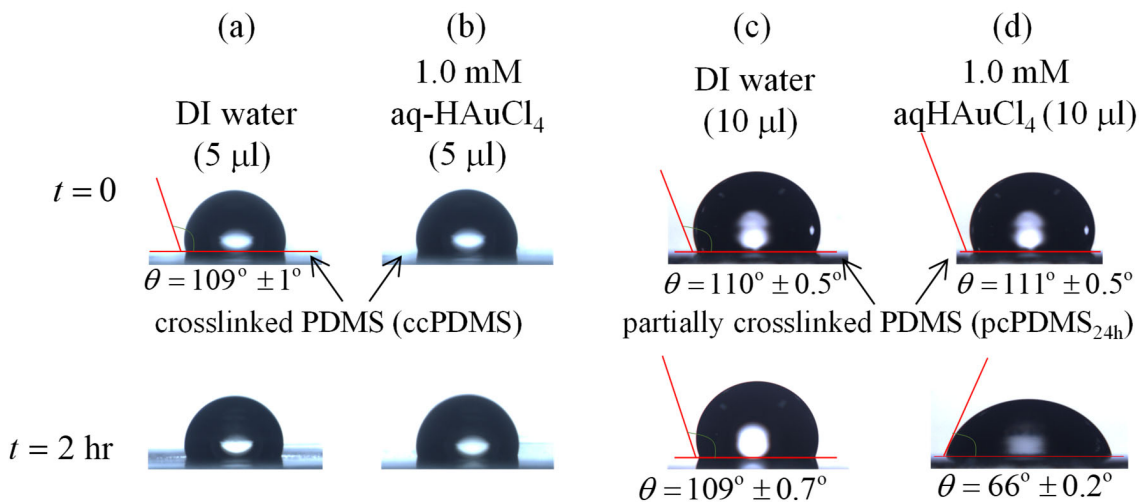

Figure S9. Drops of aq-HAuCl4 and water were gently dispensed on ccPDMS (sequence of images in column (a) and (b)) and pcPDMS$_{24h}$ (images in columns (c) and (d)) surfaces and their evolution over time was examined. At the interface of the solid and liquid, the reaction between HAuCl4 and SiH leads to the appearance of NPs which drive the contact line. Drops of HAuCl4 spread significantly on pcPDMS surfaces with decrease in advancing contact angle.

Figure S9 shows the results from experiments in which sessile drops of aqHAuCl4 ($C_{HAuCl_4} = 1.0$ mM) of volume 5-10 μl were dispensed on completely (ccPDMS) and partially (pcPDMS) crosslinked PDMS surfaces. Drops of DI water were used as control. The diameter 1.05-1.35 mm of these liquid drops was smaller than the capillary length of PDMS, i.e. 2.7 mm implying that role of gravity was insignificant on spreading of these drops. Yet, the drops were found to spread on these surfaces to different extent along with the evolution of the gold NPs (AuNPs). On a ccPDMS surface, both DI water and 1.0 mM aqHAuCl4 were found to form an advancing contact angle (CA) of $\theta_{H_2O} = \theta_{HAuCl_4} = 109° \pm 1°$. The drop of DI water didn't spread at all on this surface so that the contact angle remained unaltered over time. Even the drop of aqHAuCl4 didn't spread appreciably over 5 hours as the reaction between the R$_3$SiH molecule present in crosslinked film of PDMS and the Au-salt proceeded to a limited extent within this time.

Extent of spreading of these drops was different on the pcPDMS substrate. Here the drop of DI water did not spread any significantly; over 2 hours the macroscopic CA, $\theta_{H_2O}$ decreased from ~110° to ~109° (figure S9c), while the contact radius of the drop (volume: 10 μl) increased from 1.2 mm to 1.31 mm. In the absence of any reaction at the interface, this spreading, albeit miniscule, could be because of reconstruction of the pcPDMS surface via which hydrophilic segments present at the PDMS$_x$ surface aligns favourably with the water molecules in contact with it[8,9]. Unlike water, drops of aqHAuCl4 ($C_{HAuCl_4} = 1.0$ mM) were found to spread significantly on the pcPDMS$_{24h}$ surface (figure 4d). The contact radius of the drop increased from 1.2 mm to 1.92 mm over 2 hours; the macroscopic contact angle of the drops too decreased: from $\theta_{HAuCl_4} = 111°$ to $66°$.

**Estimation of solid-liquid interfacial energy using elasto-capillary effect in micro-channel embedded elastic film.**

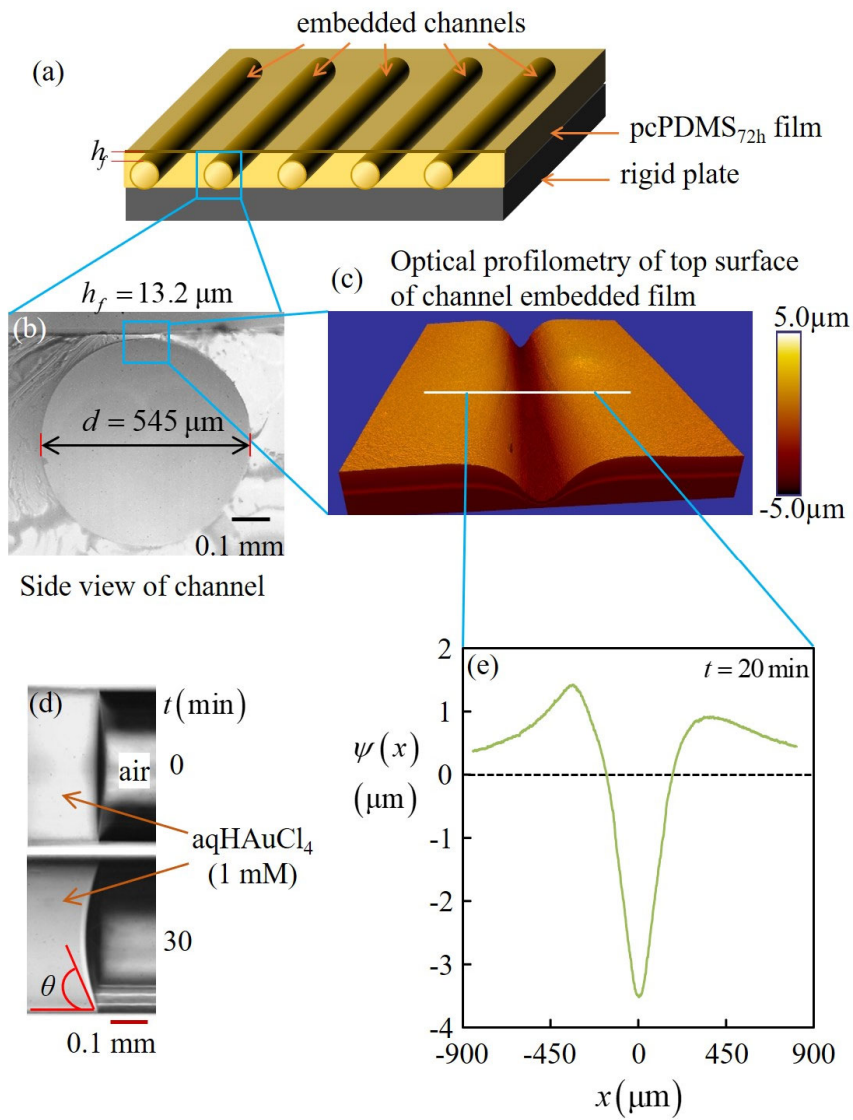

**Figure S10.** The interfacial energy of aqHAuCl$_4$ – pcPDMS$_{72h}$ films were estimated by the method depicted in reference (*10*). In brief, micro-channels of diameter, $d = 545$ μm were embedded inside thin films of pcPDMS$_{72h}$ (i.e. PDMS films crosslinked for 72 hour at room temperature ~25ºC) film bonded to a rigid substrate by the method depicted in reference (*11*) The thickness of the film was such that its minimum thickness $h_f$ at the vicinity of the **vertex of the channel was maintained at ~13 μm. aqHAuCl₄ solution ($C_{HAuCl_4} = 1.0$ mM) was inserted into the channel and the deformation of the film surface was obtained using optical profilometer. The reaction between the HAuCl₄ with SiH was allowed to occur for ~20 min following which the contact angle of the liquid meniscus inside the channel and the profile of the film surface was measured.**

**Supplementary Text S3**

**Discussion on surface stress $\Upsilon_s$ and $\Upsilon_{sl}$ at the vicinity of the sessile drop on solid:**

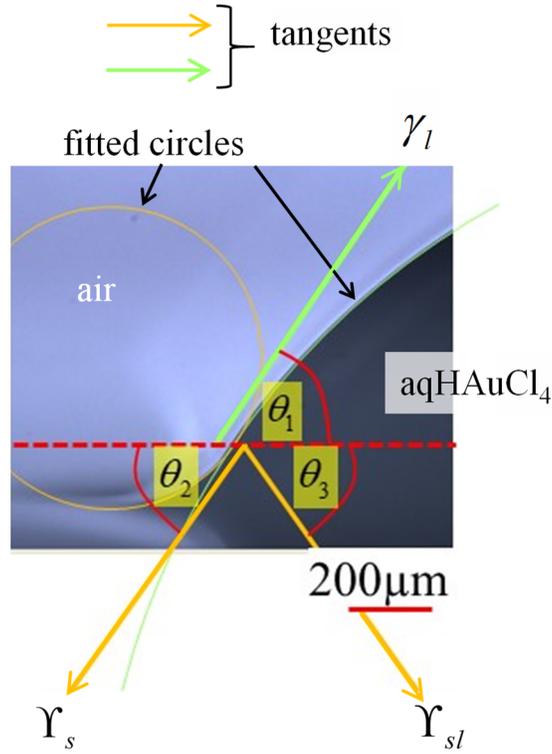

Balance of surface stresses along the $x$ and $y$ directions yield the following equations which are solved for estimating $\Upsilon_s$ and $\Upsilon_{sl}$ from the measurement of angles $\theta_1$, $\theta_2$ and $\theta_3$.

$$\gamma_{lv} \cos\theta_1 = \Upsilon_s \cos\theta_2 - \Upsilon_{sl} \cos\theta_3 \qquad 1$$

$$\gamma_{lv} \sin\theta_1 = \Upsilon_s \sin\theta_2 - \Upsilon_{sl} \sin\theta_3 \qquad 2$$

For a sessile drop of aqHAuCl4 ($C_{HAuCl_4} = 1.0$ mM) in contact with the pcPDMS24h for 2 hours, $\theta_1$, $\theta_2$ and $\theta_3$ are measured as in the table below.

| $\theta_1$ | $\theta_2$ | $\theta_3$ | $\Upsilon_s$ (mN/m) | $\Upsilon_{sl}$ (mN/m) |
|---|---|---|---|---|
| 47.7±1.5⁰ | 45.5±1.8⁰ | 45.5±1.8⁰ | 71.6±0.1 | 2.9±0.1 |

Equations 1 and 2 were solved to obtain $\Upsilon_{sl}$ and $\Upsilon_s$ had estimated values of ~2.9 mN/m and ~71.6 mN/m, respectively.

**Surface stresses $\Upsilon_s$ and $\Upsilon_{sl}$ correspond to the "current state" of the corresponding interfaces. $\Upsilon_{sl}$ for pcPDMS-water and pcPDMS-HAuCl$_4$ surfaces are expected to be different as the solid surface remains in contact with two different liquids: water and aqHAuCl$_4$. As regards the solid-air (S-A) interface, at the vicinity of the three-phase contact line of the respective liquid drops, the surfaces were not exactly identical. While for the water drop, the solid surface continued to remain non-wettable, for the drop of aqHAuCl$_4$, it turned wettable. Atomic Force Microscopy (AFM) of this surface showed that it became rough possibly because of diffusion of AuNPs from the solid-liquid to the solid-air interface (fig S11). As a result, the solid-air interface at the vicinity of the respective drops were energetically different[12].**

**Roughness of PDMS surface at the vicinity of three-phase contact line of a aqHAuCl4 drop.**

Since in conventional methods, AFM tip could not be used for scanning the soft reacted surface of pcPDMS$_{24h}$, this surface was required to be post-cured at high temperature (90º C) before scanning the surface with the AFM. However, curing resulted in diffusion of AuNPs into the PDMS network thereby rendering the AuNPs undetectable. To overcome this effect, a PDMS film of thickness ~0.7 - 0.8 mm was prepared by mixing the oligomer (Sylgard 184 elastomer) and curing agent at 10:3 weight ratio and was cured at 24º C for 38 hours. The resultant film was soft enough for the elastocapillarity induced deformation to occur at the three-phase contact line of a sessile drop of aqHAuCl4 on it, yet the film had enough quantity of SiH on its surface that allowed it to reduce the salt forming AuNPs.

In Figure S11, we show how nanoparticles form at the vicinity of the three-phase contact line of a sessile drop of 11mM aqHAuCl4 ($C_{HAuCl_4} = 11$ mM) dispensed on the surface of the above substrate and was allowed to react for 20 minutes. The drop was then gently removed and the reacted surface was cured by heating at 90 degrees for ~4 hours. This surface was subjected to optical imaging and AFM scanning.

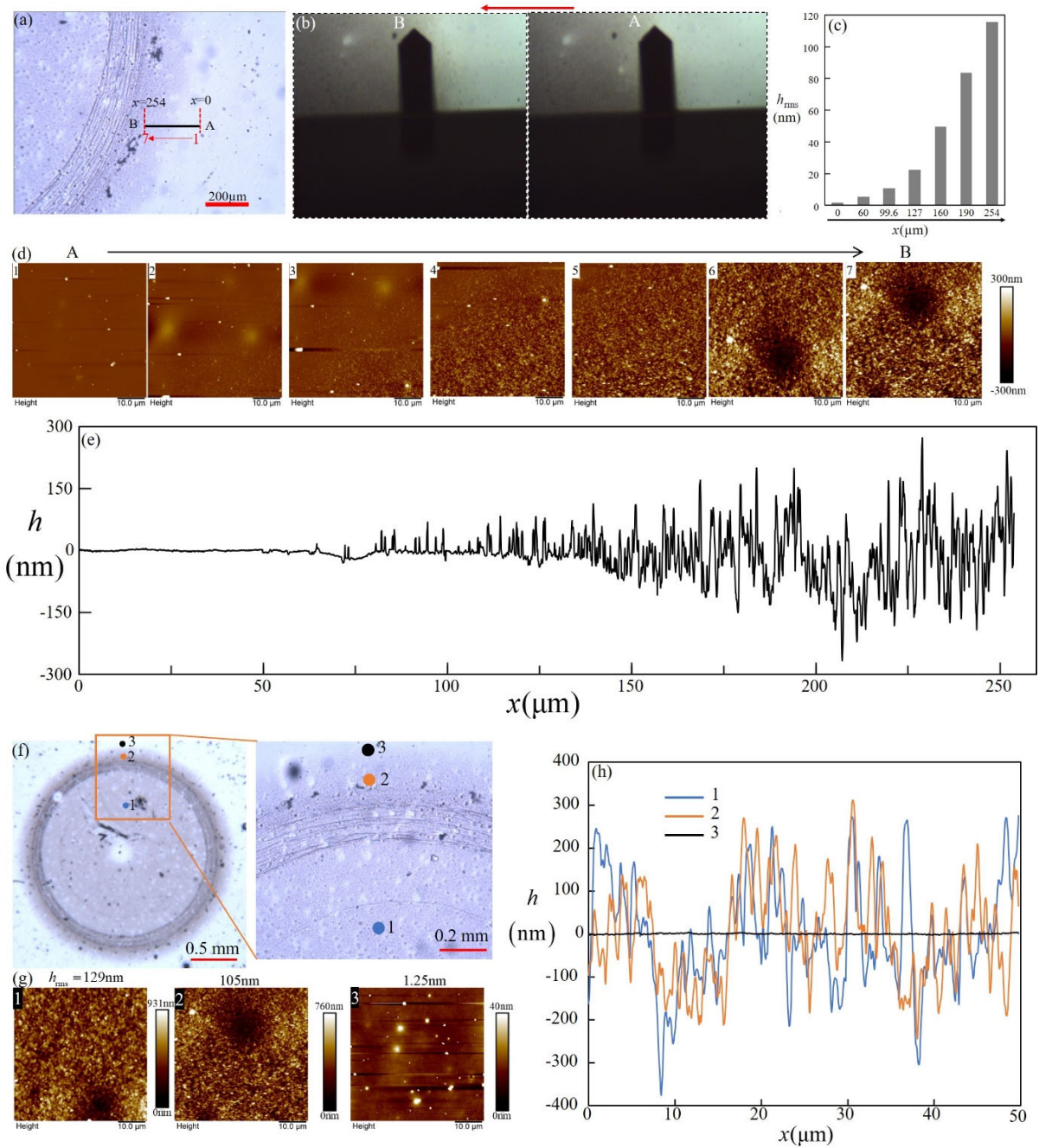

Figure S11. Surface morphology at the vicinity of three phase contact line: 10 ul drop of 11mM aqHAuCl$_4$ was placed on 38 hour crosslinked PDMS (thickness ~0.75mm, oligomer: curing agent 10:3 w/w ratio). It was allowed to react for 20 min following which the drop was first removed by rinsing with water, then dried with Nitrogen gas, following which, the film was cured completely at 90º C for 4 hour. The surface was then analyzed under optical
49

**microscope and Atomic Force Microscope (AFM). Optical micrograph (a) shows the magnified view of the surface at the vicinity of the three-phase contact line. (b) To obtain the surface morphology, an AFM cantilever was carefully moved in the x-direction in Scanasyst mode (Multimode 8, Bruker AFM) along the line AB. (c) The AFM scans at different locations along line AB. (d) The bar chart represents the root mean square roughness, $h_{rms}$ at locations 1-7 as estimated from the scanned images in figure S11(c). The increase in roughness corresponds to appearance of gold nano-particles (AuNPs). (f-g) AFM scan was done at three different locations on the surface: (1) correspond to a locations inside the wetted area of the drop; (2) corresponds to one on the three phase contact line; and location (3) represents a location far away from the contact line. The roughness of the surface are similar at locations 1 and 2 and decreases at 3.**

**Supplementary text S4**

**Estimation of excess free energy per mole of the salt solution, $\overline{W}$.**

To obtain $\overline{W}$, dilute solution limit of the salt was considered, at which, the aqueous solution was assumed to consist essentially of water molecules. Considering that water molecules are spherical, a two dimensional area that can be covered by one mole of the water molecules was deduced[13], $A_m = (N_A)^{1/3} (V_m)^{2/3}$, where $V_m$ is the molar volume and $N_A = 6.023 \times 10^{23}$ is the Avogadro number. For water, putting $V_m = 18$ cc/mole, $A_m$ can be estimated as $58 \times 10^3$ m²/mole. Thereafter, relating the excess free energy per mole, $\overline{W}$ to that per unit area, $W$ by the relation: $\overline{W} = W \cdot A_m$, for aqHAuCl4 in contact with pcPDMS$_{24h}$ ($W_{HAuCl_4}^{pool} \sim 91$ **mJ/m²**, $W_{HAuCl_4}^{drop} \sim 140.5$ **mJ/m²**), $\overline{W}$ for the pool and drop geometry, was estimated as $\overline{W}_{HAuCl_4}^{pool} \sim 5.28$ kJ/mole and $\overline{W}_{HAuCl_4}^{drop} \sim 8.16$ kJ/mole respectively.